\documentstyle[aps,preprint]{revtex}

\def\centereps#1#2#3{\vskip#2\relax\centerline{\hbox to#1{\special
  {eps:#3 x=#1, y=#2}\hfil}}}

\widetext
\draft
\tighten
\def\abstract{\if@twocolumn
\section*{Resumen}
\else \small
\begin{center}
{\bf Resumen\vspace{-.5em}\vspace{0pt}}
\end{center}
\quotation
\fi}

\def\thebibliography#1{\section*{Referencias\markboth
{REFERENCIAS}{REFERENCIAS}}
\list
{[\arabic{enumi}]}{\settowidth\labelwidth{[#1]}
\leftmargin\labelwidth
\advance\leftmargin\labelsep
\usecounter{enumi}}
\def\newblock{\hskip .11em plus .33em minus -.07em}
\sloppy
\sfcode`\.=1000\relax}

\begin{document}

\preprint{EFUAZ FT-96-22}

\title{De Dirac a Maxwell: un Camino con Grupo de
Lorentz\thanks{Presentado en el Primer Encuentro de Investigaci\'on,
Zacatecas, UAZ, 28 de Febrero al 1 de Marzo, 1996. Enviado a
``Investigaci\'on Cientifica".}}

\author{{\bf Valeri V. Dvoeglazov}}

\address{
Escuela de F\'{\i}sica, Universidad Aut\'onoma de Zacatecas \\
Antonio Doval\'{\i} Jaime\, s/n, Zacatecas 98068, ZAC., M\'exico\\
Correo electronico:  VALERI@CANTERA.REDUAZ.MX}

\date{30 de Abril de 1996}

\maketitle

\abstract{
Es una introducci\'on en nivel accesible a recientes ideas dirigidas
para unificar las interacciones en base del grupo de Lorentz.
Se demuestra que tanto la ecuaci\'on de Dirac como las ecuaciones
de Maxwell podrian ser consideradas como consequencia de los
mismos postulados. Se discuten probables ge\-nera\-liza\-ciones de
las ecuaciones para part\'{\i}culas del espin 1 (fotones)
en area de altas frecuencias.}


\newpage

\medskip
\begin{flushright}
``{\it Mathematics loves symmetries above all"}\\
James C. Maxwell \hspace*{7mm}
\end{flushright}
\smallskip

\setlength{\baselineskip}{24pt}

\section{Introducci\'on}

Antes de presentar el formalismo matem\'atico y conclusiones f\'{\i}sicas
deducidas en base de las recientes ideas en mec\'anica cu\'antica
relativista, yo quisiera atraer su atenci\'on al titulo de mi
contribuci\'on. De acuerdo con la linea hist\'orica del desarollo de la
ciencia f\'{\i}sica los nombres de los famosos f\'{\i}sicos, que se
encuentran en el t\'{\i}tulo, tienen que estar en la siguiente
secuencia:  James C.  Maxwell (1831 - 1879), Hendrik A.  Lorentz (1853
- 1928), Paul A.  M.  Dirac (1902 - 1984).  Entonces, ?` por qu\'e de
Dirac a Maxwell? En mi opini\'on la linea l\'ogica del desarollo de la
ciencia no siempre coincide con la linea hist\'orica, cronolog\'{\i}a.
Siempre existieron unos cient\'{\i}ficos que adelantaban su tiempo; siempre
los tiempos de revoluciones en la ciencia cambiaron por los periodos del
conservatismo y viceversa; siempre una de las tareas de la ciencia fue
servir a la tecnolog\'{\i}a y resolver los problemas del presente; pero
siempre, tamb\'{\i}en, los pensadores intentaban unificar las ideas y
conceptos para facilitar la descripci\'on de la naturaleza y
la aplicaci\'on practica podra ser notada \'unicamente en futuro.  Desde
este punto de vista, como convenderemos despues, es mejor para la busqueda
de la teor\'{\i}a unificada de las interacciones f\'{\i}sicas echar una
mirada a los formalismos~\cite{Maxwell,Lorentz,Dirac}, que se usan por
ahora, en el orden inverso con respeto al orden cronol\'ogico.

La base de la f\'{\i}sica moderna es la teor\'{\i}a de la relatividad y
teor\'{\i}a cu\'antica. La mayor parte de ellas fu\'e creada por
Albert Einstein (1879 - 1955),
el rey  de la raza de `los gigantes' que reformular\'on las leyes
clasicas de la naturaleza. ?`Qui\'en no sabe su famosa relaci\'on
entre energ\'{\i}a $E$ y masa $m$, ref.~\cite{Einstein}?  $${\bbox
E=mc^2}$$
Pero \'el dijo acerca de si mismo, ``yo hice algo en la ciencia,
unicamente porque estaba en los hombros de los gigantes", implicando
tan grandes f\'{\i}sicos y matem\'aticos como Von Helmholtz, Maxwell,
Lorentz, Poincar\`e, Hilbert y muchos otros. Su sue\~no fue crear una
teor\'{\i}a que unificara las interacciones gravitacionales y
electromagn\'eticas, las que eran conocidas en aquellos tiempos, y las
ideas cu\'anticas.  En~\cite[p. 299]{Stephani} es mencionado:  {\it ``For
a time he [A. Einstein] believed that quantum theory could be encompassed
in a (possibly generalized) theory of relativity that would link
space-time singularities to elementary particles. This hope has not been
realized"\ldots} Ademas, desafortunadamente, este `marriage' entre
teor\'{\i}a de relatividad restringida (o, tambien, generalizada) y
mec\'anica cu\'antica de Heisenberg y Schr\"odinger no habia sido realizado
por completo hasta los presentes d\'{\i}as a pesar de muchos
que lo intentaron.

En un dibujo de Bulent Atalay (1978) dos portretos  (Albert Einstein y
Eugene Wigner (1902 - 1995)). Einstein mira a Wigner, el matematico y el
f\'{\i}sico que compus\'o~\cite{Wigner} la teor\'{\i}a de las
representaciones del grupo de Lorentz inhomogenio, que se conecta con
las propiedades intrinsecas de nuestro espacio y tiempo y que se usa
en todas las teor\'{\i}as que tienen invariancia relativista.
Tal vez, Einstein no realiz\'o  su sue\~no en realidad porque no estaba en
los hombros de Wigner ?`Qui\'en sabe?

\section{Los grupos de Lorentz y Poincar\`e. Sus representaciones}

De acuerdo con el principio de relatividad las leyes
f\'{\i}sicas son invariantes  con respecto del cambio de un sistema de
referencia. En lengua matem\'atica esto puede expresarse  como
la transformaci\'on lineal de coordenadas:
\begin{equation}
x^\mu \rightarrow x^{\prime^{\mu}} = L^\mu_{\quad\nu} \, \, x^\nu +a^\mu
\equiv L^{\mu\nu} x_\nu + a^\mu\quad,\quad L^{\mu\nu}=
-L^{\nu\mu}\quad,
\end{equation}
con condici\'on de conservaci\'on del intervalo $s$. La forma diferencial
de lo \'ultimo es la siguiente:
\begin{equation}
ds^2 = dx_0^2 - d{\bf x}^2\quad,\quad x_0 =ct\quad,
\end{equation}
$c$ es la velocidad de la luz, $t$ es la coordenada de tiempo y
${\bf x}$ son las coordenadas espaciales, $x^\mu = (x_0, {\bf x})$. Se
puede parametrizar esas transformaciones en t\'erminos de los \'angulos de
Euler y la velocidad de un sistema de referencia con respeto de  otro
(vease, por ejemplo,~\cite{Ryder}).  Entonces, podemos dar la siguiente
definici\'on:  {\it El grupo de Lorentz propio (ortochroneo) es el grupo
de rotaciones en los planos $XY, XZ, YZ$ y  transformaciones puras de
Lorentz (} {\tt boosts}, {\it las rotaciones en planos pseudoeuclideanos
$XT, YT, ZT$). El grupo de Poincar\`e incluye tambi\'en translaciones del
punto central del sistema  de referencia en espacio-tiempo. Esas
transformaciones guardan intervalo entre eventos.} El grupo de Lorentz se
define por seis par\'ametros, el grupo de Poincar\`e, por diez. Se
consideran tambien los grupos extendidos que incluyen reflexiones de los
ejes de espacio (${\bf x}\rightarrow -{\bf x}$) y de tiempo
($x_0=ct\rightarrow -x_0$).\footnote{Generalmente, en teor\'{\i}a
cu\'antica de campos se usa el sistema de unidades en cual $c=1$ (por
conveniencia). En el resto del articulo vamos a usar frequentamente esa
sistema.}

En las teor\'{\i}as cu\'anticas las part\'{\i}culas se describen por
la funci\'on de onda (o, bien, por la funci\'on del campo) que depende de
cuatro coordenadas $x^\mu$. La funci\'on del campo puede ser, en su
turno, la funci\'on con unas componentes. Por ejemplo, para la
descripci\'on de  un electr\'on se usa la funci\'on con cuatro componentes,
{\tt bispinor};\footnote{En teor\'{\i}as que describen cambios del n\'umero
de part\'{\i}culas en alg\'un proceso es necessario proyectarla en el
espacio de Fock, de hecho, un espacio de infinita dimensi\'on. En otras
palabras es necessario introducir las {\tt operadores de creaci\'on y
aniquilaci\'on}.} para un fot\'on, tambi\'en con cuatro componentes, el
4-vector potencial o, bien, se puede usar la funci\'on con seis
componentes que corresponden a los componentes del campo el\'ectrico y el
campo magn\'etico, los observables de el\'ectrodinamica clasica. Entonces,
la diferencia entre electr\'on y fot\'on es que sus funciones de
campo transforman de acuerdo con diferentes representaciones del grupo de
Poincar\`e.  Matematicamente, a un cambio de sistema de referencia se puede
poner en correspondencia una transformaci\'on lineal y uniforme de las
funciones del campo (espinorial, o escalar, o vectorial, o algun otro)
\begin{equation}
u (x) \Longrightarrow u^\prime (x^\prime) = \Lambda u
(x)\quad.
\end{equation}
La matriz $\Lambda$ se define completamente por
la matriz $L$ y al producto de dos elementos del grupo de Poincar\`e ponen
en correspondencia el producto de dos transformaciones de las funciones
del campo:
\begin{equation}
\Lambda_{L_1 L_2} = \Lambda_{L_1}
\Lambda_{L_2}\quad.
\end{equation}
Al elemento unidad $I$ de grupo de
Poincar\`e  corresponde la transformaci\'on id\'entica de las funciones, al
elemento inverso corresponde la transformaci\'on inversa en espacio de las
funciones del campo. Este es una {\tt representaci\'{o}n} del grupo de
Poincar\`{e};  $u(x)$ forman las  funciones de base de la
transformaci\'on.  Se puede definir $({1\over 2},0)\oplus (0,{1\over 2})$,
$(0,0)$, $({1\over 2},{1\over 2})$, $(1,0)\oplus (0,1)$ etc.
representaciones. Los n\'umeros indicados tiene sentido de los valores
propios del operador de momento angular para ambas partes de la funci\'on
del campo.

\section{Ecuaci\'on de Dirac: dos caminos para deducir}

P. A. M. Dirac  dedujo su famosa ecuaci\'on para descripci\'on de un
electr\'on y su antipart\'{\i}cula, {\tt positron} en 1928. A esa fecha
ya fueron formados y aceptados ambas teor\'{\i}a de relatividad y
mec\'anica cu\'antica. El problema fue proponer la ecuaci\'on que
toma en cuenta los efectos relativistas y que describe bien las
part\'{\i}culas conocidas en aquellos tiempos.   La ecuaci\'on de
Schr\"odinger que ya existia
\begin{equation}
i\hbar {\partial \Psi ({\bf x}, t) \over \partial t} =
\hat H \Psi ({\bf x}, t)
\end{equation}
no fue relativista. La cuenta de correciones relativistas (por ejemplo,
la dependencia de la masa con velocidad) fue muy dificil.
La ecuaci\'on de Klein y Gordon
\begin{equation}
\left [ \nabla^2 - {1\over c^2} {\partial^2 \over \partial t^2}
\right ] \Psi ({\bf x}, t) = m^2 \Psi ({\bf x}, t)\quad,
\end{equation}
aunque fue relativista, pero tenia gran defecto: la norma (de hecho,
la probablidad de acuerdo con interpretaci\'on de mecanica cu\'antica
por N.  Bohr) no fue la cantidad definida a ser positiva.
Dirac propuso que la ecuaci\'on correcta tiene que
satisfacer a la relaci\'on dispercional relativista de Einstein
(como la ecuaci\'on de Klein y Gordon la satisface)
\begin{equation}
E^2 = {\bf p}^2 c^2 +m^2 c^4 \quad,
\end{equation}
y, ademas, de ser de primer orden en las derivadas en tiempo y en
coordenadas espaciales
\begin{equation}\label{ham}
i\hbar {\partial \Psi (x^\mu) \over \partial t} = \left [\alpha_1 {\partial
\over \partial x} + \alpha_2 {\partial \over \partial y} +\alpha_3
{\partial \over \partial z} +\beta m\right ] \Psi (x^\mu)\quad.
\end{equation}
El lleg\'o al resultado: para cumplir estos requisitos
era necesario suponer:
\begin{itemize}
\item
Que $\alpha_i$ y $\beta$ sean matrices que satisfacen las relaciones
de comutaci\'on:
\begin{mathletters}
\begin{eqnarray}
\left \{\alpha_i , \alpha_k \right \}_+ &=& 0\quad \mbox{si} \quad i \neq
k\quad,\\
\left \{\alpha_i , \,\beta\,\, \right \}_+ &=& 0\quad,\quad  \alpha_i^2
=\beta^2 =\openone\quad.
\end{eqnarray}
\end{mathletters}
Rango de esas matrices tiene
que ser no menos que  cuatro (la dimensi\'on minima es $4\times 4$).

\item
Existencia de antiparticulas, teor\'{\i}a de {\tt huecos}
y el {\tt `mar de Dirac'}. El \'ultimo, de hecho, expone estructura
complicada de vacio, este es el mar infinito de electrones, protones y
otras part\'{\i}culas con energia negativa y con el espin $1/2$. Si un
`electr\'on' de este mar soporta una transici\'on al area de energias
positivas (v\'ease el grafico 1) por la absorbci\'on de energ\'{\i}a, una
vacante resultante en este mar puede considerarse como una
part\'{\i}cula, positron, la antipart\'{\i}cula de un electron. El proceso
inverso (aniquilaci\'on) puede occurirse unicamente cuando tenemos una
vacante (un `hueco') en el area de energias negativas. La transici\'on
sin hueco alla es prohibido por el principio de Pauli.

\end{itemize}

\vspace*{1mm}

La forma  (\ref{ham}) se llama la forma hamiltoniana, pero por 
la redefinici\'on de las matrices la ecuaci\'on de
Dirac se reescribiria a la forma covariante ($\partial_\mu \equiv
{\partial \over \partial x^\mu}$)
\begin{equation}\label{cf}
\left [i\gamma^\mu
\partial_\mu -m \right ] \Psi (x^\mu) = 0\quad.
\end{equation}
Existe el otro camino para deducir la ecuaci\'on de Dirac.
Este es por uso de los postulados de Wigner. Para construcci\'on
de las teor\'{\i}as relativistas en cualquier representaci\'on
vamos suponer que:

\begin{itemize}

\item
Relaciones dispersionales relativistas $E^2 -{\bf p}^2 =m^2$,\,
son validos para estados de part\'{\i}culas libres.

\item
Para $j$ espin arbitrario los derechos $(j,0)$ y los izquierdas
$(0,j)$ {\tt espinores}  transforman de acuerdo con las reglas de
Wigner:
\begin{eqnarray}
\phi_{_R} (p^\mu) &=& \Lambda_{_R} (p^\mu \leftarrow
\overcirc{p}^\mu ) \phi_{_R} (\overcirc{p}^\mu) = \exp (+ {\bf J}\cdot
{\bbox \varphi})\phi_{_R} (\overcirc{p}^\mu)\quad,\\
\phi_{_L} (p^\mu) &=&
\Lambda_{_L} (p^\mu \leftarrow \overcirc{p}^\mu ) \phi_{_L}
(\overcirc{p}^\mu) = \exp (- {\bf J}\cdot {\bbox \varphi})\phi_{_L}
(\overcirc{p}^\mu)\quad.
\end{eqnarray}
$\Lambda_{_{R,L}}$ son matrices de {\tt
boost} de Lorentz; {\bf J} son matrices de espin; ${\bbox \varphi}$ son
parametros de {\tt boost} dado. En caso de particulas {\tt bradyons} los
\'ultimos se definen:\footnote{Los {\tt bradyones} son part\'{\i}culas
que se mueven con la velocidad menos que la velocidad de la luz
$c\approx 300,000$ km/s. Los {\tt taquiones}, con la velocidad  mayor
que $c$.  La idea de existencia de
taquiones fue desarollado por Prof.  E.  Recami hace mucho tiempo,
pero solo en los \'ultimos a\~nos fueron producidos los
experimentos que indican la existencia de los paquetes de onda
{\tt taquionicos}, {\tt the $X$ waves}. Esperamos que los resultados
sean confirmados.}
\begin{equation}
\cosh (\varphi) = \gamma =
\frac{1}{\sqrt{1-v^2}} = {E\over m}\quad,\quad \sinh (\varphi) = v\gamma =
{\vert {\bf p} \vert \over m}\quad.
\end{equation}
$\hat {\bbox \varphi}
= {\bf n} = {\bf p}/\vert {\bf p}\vert$
si trabajamos en sistema de
unidades $c=1$.

\item
$\phi_{_{R,L}} (p^\mu)$ por transformaci\'on unitaria puede
arreglar ser los {\tt espinores} propios del operador de {\tt helicidad}:
\begin{equation}
({\bf J}\cdot {\bf n}) \phi_{_{R,L}} = h \phi_{_{R,L}}\quad,\quad
h=-j,-j+1,\ldots j\quad.
\end{equation}

\item
En reposo (exactamente, si escogemos el sistema de referencia en tal manera
que $\overcirc{p}^\mu = (E=m, {\bf p} = {\bf 0})$; no existen
part\'{\i}culas sin masa en reposo) {\tt espinores} derechos e izquierdas
se conectan
\begin{equation} \phi_{_R} (\overcirc{p}^\mu) = \pm \phi_{_L}
(\overcirc{p}^\mu) \quad.
\end{equation}
Esta es relaci\'on que se llama la
relaci\'on de Ryder-Burgard~\cite{Ryder}.
\end{itemize}

Vamos asentar ${\bf J} = {\bbox \sigma}/2$ y formar un objeto
cuatridimensional:
\begin{equation}
\Psi (p^\mu) = \pmatrix{\phi_{_R} (p^\mu)\cr
\phi_{_L} (p^\mu)\cr}\quad.
\end{equation}
Despues de applicaci\'on de las reglas indicados y la relaci\'on de Ryder
y Burgard obtenemos
\begin{mathletters}
\begin{eqnarray}
\phi_{_R} (p^\mu) &=& \Lambda_{_R} (p^\mu \leftarrow \overcirc{p}^\mu)
\phi_{_R} (\overcirc{p}^\mu) =\\
&=&\pm \Lambda_{_R} (p^\mu
\leftarrow \overcirc{p}^\mu ) \phi_{_L} (\overcirc{p}^\mu) = \pm
\Lambda_{_R} (p^\mu \leftarrow \overcirc{p}^\mu) \Lambda_{_L}^{-1} (p^\mu
\leftarrow \overcirc{p}^\mu) \phi_{_{L}} (p^\mu)\quad,\nonumber\\
\phi_{_L} (p^\mu)
&=& \Lambda_{_L} (p^\mu \leftarrow \overcirc{p}^\mu) \phi_{_L}
(\overcirc{p}^\mu) = \\
&=&\pm \Lambda_{_L} (p^\mu \leftarrow
\overcirc{p}^\mu) \phi_{_{R}} (\overcirc{p}^\mu) = \pm \Lambda_{_L} (p^\mu
\leftarrow \overcirc{p}^\mu) \Lambda_{_R}^{-1} (p^\mu \leftarrow
\overcirc{p}^\mu) \phi_{_{R}} (p^\mu)\quad.\nonumber
\end{eqnarray}
\end{mathletters}
Desarrollando
\begin{equation} \exp(\pm{\bbox\sigma}
\cdot {\bbox\varphi}/2) = \openone \cosh {\varphi\over 2} \pm
({\bbox\sigma}\cdot \hat {\bbox\varphi}) \sinh{\varphi\over 2}
\end{equation}
y usando que para las representaciones del grupo de Lorentz
finito-dimensionales tenemos $\Lambda_{_{L,R}}^{-1} =
\Lambda_{_{R,L}}^\dagger$
llegamos a la forma de matriz de la ecuaci\'on
de Dirac en espacio del momento lineal:
\begin{eqnarray}
\pmatrix{\mp m\openone & E+({\bbox
\sigma}\cdot {\bbox p})\cr E-({\bbox \sigma}\cdot {\bbox p})& \mp m
\openone\cr}\Psi (p^\mu) =0 \quad.
\end{eqnarray}
Introduciendo $\Psi (x^\mu)
\equiv \Psi (p^\mu) \exp (\mp ip_\mu x^\mu)$ immediatamente recibemos la
forma covariante en espacio de coordenadas, la ecuaci\'on (\ref{cf}).

\section{Ecuaciones de Maxwell (1864) y ecuaciones de~Weinberg~(1964)}

S. Weinberg\footnote{El es laureato del Premio de Nobel, esta
trabajando ahora en Universidad de Texas en Austin, EUA} estaba
considerando las siguientes ecuaciones para el campo con espin $j$,
ref.~[8b,p.B888]:
\begin{mathletters} \begin{eqnarray}\label{w1} \left [
{\bf J}^{(j)} \cdot \bbox{\nabla} - j (\partial / \partial t) \right ]
\varphi (x) &=& 0\quad,\\ \label{w2} \left [ {\bf J}^{(j)} \cdot
\bbox{\nabla} + j (\partial / \partial t) \right ] \chi (x) &=& 0\quad.
\end{eqnarray} \end{mathletters}
El declar\'o:  {\it `` For $j=\frac{1}{2}$
these are the Weyl equations for the left- and right-handed neutrino
fields, while for $j=1$ they are just Maxwell's free-space equations for
left- and right-circularly polarized radiation:
\begin{mathletters}
\begin{eqnarray}
\bbox{\nabla} \times  [ {\bf E} - i{\bf B}] +
i(\partial/\partial t) [{\bf E} - i{\bf B}] &=& 0\quad,\quad\qquad
(4.21)\nonumber\\
\bbox{\nabla} \times [ {\bf E} + i{\bf B}] -
i(\partial/\partial t) [{\bf E} + i{\bf B}] &=& 0\quad,\quad\qquad
(4.22)\nonumber \end{eqnarray} \end{mathletters}
The fact that these field
equations are of first order for any spin seems to me to be of no great
significance\ldots}"(!) ?`Por qu\'e el lleg\'o a tan sorprendente
conclusi\'on?  ?`Y por qu\'e los f\'{\i}sicos no pagaron ninguna
atenci\'on a esa idea hasta los a\~nos noventas?

Vamos a considerar un poco las ideas basicas de su teor\'{\i}a.
Si seguimos en misma manera a la deducci\'on que usamos antes (para la
ecuaci\'on de Dirac, $j=1/2$), esta vez con applicaci\'on al caso
del espin 1 podemos obtener la siguiente ecuaci\'on~\cite{BWW}
\begin{equation}
\left [ \gamma^{\mu\nu} \partial_\mu \partial_\nu
+\wp_{u,v} m^2 \right ] \Psi (x^\mu) =0\quad.
\end{equation}
Esta es la
ecuaci\'on en el $(1,0)\oplus (0,1)$ representaci\'on, la representaci\'on
bivectorial.  $\wp_{u,v}=\pm 1$ depende de cual soluci\'on, con
energ\'{\i}a positiva o negativa, se considere.  Una de las consequencias
f\'{\i}sicas importantes es el hecho que teoricamente los {\tt
bosones}, part\'{\i}culas con el espin entero, pueden llevar el numero
cu\'antico, la {\tt paridad}, opuesto con respecto de su
antipart\'{\i}cula.  Este es el ejemplo explicito de las teor\'{\i}as
predicados por Bargamann, Wightman y Wigner~[6b].  Luego, en el limite sin
masa  $m=0$ y $g_{\mu\nu} p^\mu p^\nu \phi_{_{R,L}} (p^\mu) = 0 = p^2
\phi_{_{R,L}} (p^\mu)$ podemos obtener las ecuaciones
\begin{mathletters}
\begin{eqnarray} \label{a1}
2 ({\bf J}\cdot {\bf p})
\left [ ({\bf J}\cdot {\bf p}) +E \openone\right ] \phi_{_R}
(p^\mu)&=&0\quad, \\ \label{a2} 2 ({\bf J}\cdot {\bf p}) \left [ ({\bf
J}\cdot {\bf p}) -E \openone\right ] \phi_{_L} (p^\mu)&=&0\quad.
\end{eqnarray}
\end{mathletters}
Podemos ver que ellas se difieren de las ecuaciones
(\ref{w1},\ref{w2}).  Los \'ultimos, en caso de selecci\'on de la base
de los operadores del espin como $({\bf J}_i)_{jk} = -i\epsilon_{ijk}$,
$\epsilon_{ijk}$ es el tensor de Levi-Civita, son la forma de
Majorana-Oppenheimer~\cite{Majorana,Oppenheimer} de las ecuaciones de
Maxwell, como denot\'o Weinberg (4.21,4.22).  $\phi_{_{R,L}}$ son
``espinores" en el $(1,0)\oplus (0,1)$ representaci\'on y tiene que ser
interpretados como {\tt bivectores} $\phi_{_{R,L}} \equiv {\bf E} \pm i
{\bf B}$.  Pero, las ecuaciones (\ref{a1},\ref{a2}) son del segundo orden
en derivadas.

Entonces, podemos observar las siguientes consecuencias de diferentes
formas de las ecuaciones en $(1,0)\oplus (0,1)$ representaci\'on.

\begin{enumerate}
\item
Las ecuaciones de Weinberg y de Maxwell (en forma de Majorana
y Oppenheimer) son muy parecidas. Pero hay diferencias: 1) WE tienen
soluciones con $E=\pm p$, 2) ME tiene soluciones con $E=\pm p$ y $E=0$. El
origen matem\'atico de esa diferencia es: la matriz $({\bf J}\cdot {\bf
p})$ no es invertible.  Este hecho fue observado en unos art\'{\i}culos en
los \'ultimos sesenta a\~nos, por \'ultima vez en los
trabajos~\cite{DVA}.\footnote{De hecho, esas articulos de Ahluwalia y
co-autores empiezan el ciclo de otros trabajos de unos grupos de
investigadores en todo el mundo.} !`No podemos ignorar la soluci\'on con
energ\'{\i}a zero!  Esa soluci\'on es para cantidades observables, los
campos el\'ectricos y magn\'eticos.

\item
En mi trabajo reciente~\cite{DVO4} y los trabajos antecedentes fueron
analizados invariantes dinamicos en base del formalismo de Lagrange y,
particularmente, el limite sin masa. Fueron obtenidas las siguientes
ecuaciones:
\begin{mathletters}
\begin{eqnarray} \label{n1}
{1\over c^2}
\frac{\partial {\bf J}_e}{\partial t} &=& - grad \,\tilde \rho_e \quad,\\
\label{n2}
{1\over c^2} \frac{\partial {\bf J}_m}{\partial t} &=& -grad \, \tilde
\rho_m \quad,
\end{eqnarray}
\end{mathletters}
(sistema de unidades electromagn\'eticos) que completan las ecuaciones de
Maxwell:
\begin{mathletters}
\begin{eqnarray}
\tilde \rho_e &=& {1\over 4\pi c^2} div {\bf E}\quad,\\
\tilde \rho_m &=& {1\over 4\pi} div {\bf B}\quad,\\
{\bf J}_e &=& {1\over 4\pi} rot {\bf B} -{1\over 4\pi c^2} \frac{\partial
{\bf E}}{\partial t}\quad,\\
{\bf J}_m &=& - {1\over 4\pi} rot {\bf E} -{1\over 4\pi}
\frac{\partial {\bf B}}{\partial t}\quad.
\end{eqnarray}
\end{mathletters}
En el caso particular, cuando se consideran unicamente
los modos transversales, se puede usar unicamente los ultimos.
En caso de existencia de la masa del fot\'on las ecuaciones
(\ref{n1},\ref{n2}) podrian ser modificados para incluir  los terminos
$m^2 {\bf E}$ and/or $m^2 {\bf B}$.

En esa formulaci\'on mas conveniente considerar que
los corrientes y cargas se definen por los campos (opuesto a las
teor\'{\i}as convencionales).
Probablemente, lo  implica que la carga y el corriente
principalmente son cantidades {\bf no-locales}. Esa idea es
reformulaci\'on de la idea vieja acerca de posibilidades de existencia de
`acci\'on en la distancia', la que discute en los trabajos recientes del
doctor A.  Chubykalo~\cite{Chubykalo} que va a presentar su platica despues
que yo.

\item
La formulaci\'on en base de ideas de Weinberg y Ahluwalia nos permite
explicar una paradoja:  hace mucho tiempo~\cite{KR} fue
declarado que el campo antisimetrico tensorial sea `longitudinal' (!?)
despues de cuantizaci\'on. Este es la violaci\'on del principio de
correspondencia, uno de mas grandes principios f\'{\i}sicos en el siglo
XX. Deducimos que esa paradoja aparece como la consequencia de
la aplicaci\'on de la condici\'on de Lorentz generalizada $\partial_\mu
F^{\mu\nu} (x^\alpha) =0$ a los estados cu\'anticos, lo que, como vemos
de las ecuaciones (\ref{a1},\ref{a2}) o de la
existencia de las ecuaciones (\ref{n1},\ref{n2}), no tiene suficiente
justificaci\'on.  En realidad, si consideramos $F^{\mu\nu} (x^\alpha)$
como un operador de campo, el vector de Pauli y Lyuban'sky ${\bf W}/E_p
={\bf J}$, que nos da valores propios del espin, es igual:
\begin{mathletters} \begin{eqnarray}
&& (W_\mu \cdot n^\mu) = - ({\bf
W}\cdot {\bf n}) = -{1\over 2} \epsilon^{ijk} n^k J^{ij} p^0\quad,\\ &&
{\bf J}^k = {1\over 2} \epsilon^{ijk} J^{ij} = \epsilon^{ijk} \int d^3
{\bf x} \left [ F^{0i} (\partial_\mu F^{\mu j}) + F_\mu^{\,\,\,\,j}
(\partial^0 F^{\mu i} +\partial^\mu F^{i0} +\partial^i F^{0\mu} ) \right
]\quad.
\end{eqnarray} \end{mathletters}
Comparando esos resultados con
el reciente  trabajo de Evans~\cite{Evans1} podemos revelar que la
formulaci\'on de Weinberg, desarollada por Ahluwalia y por mi, es
conectada con el nuevo concepto de modos longitudinales de
electromagn\'etismo, el {\tt $B(3)$ campo}~\cite{Evans} del Prof.  M.
Evans y el gran J.-P. Vigier, colaborador del fundador de mec\'anica
cu\'antica y teor\'{\i}a de la onda-pilota, el gran L. de Broglie.
Quisiera mencionar que estas teor\'{\i}as todavia estan desarollando y,
como podemos ver, son conectados con el concepto del espin, el adicional
variable sin fase, predicado por Wigner~\cite{Wigner}.

\item
Para entender que es un fot\'on, necesario crear una teor\'{\i}a
cu\'antica  en base de las ideas de Majorana~\cite{Majoran1}, construir
los estados auto/contr-auto conjugados de carga el\'ectrica. Ese trabajo
fue empezado en~\cite{NP}. Como todas las teor\'{\i}as relativistas
invariantes nuestros modelos satisfacen a los postulados que fueron
presentados en la seccion antecedente.

\item
Probablemente, mis ideas m\'as recientes, asi como de otros grupos,
esten conectados con astrof\'{\i}sica. Desde los sesentas hasta su
muerte el Prof. M. A. Markov defend\'{\i}a la idea
de friedmones, v\'ease, por ejemplo, ref.~\cite{Markov}. El declaro que
puede existir los universos (por ejemplo, nuestro Universo) que son
semi-cerrados y se puede viajar en otro Universo. En lengua del
formalismo discutido, esto significa viajar ``a traves del espejo", en un
mundo de {\tt taquiones}.  Ademas, en 1994 Prof. A. Linde~\cite{Linde}
propuso la idea que nuestro Universo es un gran monopolo.\footnote{La
primera idea de monopolos, los cuantos de la carga magn\'etica, fue
propuesta por Dirac en los treintas.}  Todos esas ideas requieren
verificaci\'on en base del formalismo presentado.

\end{enumerate}

Finalmente, como conclusi\'on, yo quiero citar unas palabras del libro de
muy conocido f\'{\i}sico de ultimos a\~nos, Asim Barut~\cite{Barut}:{\it
``Electrodynamics and the classical theory of fields remain very much
alive and continue to be the source of inspiration for much of the modern
research work in new physical theories..." }.

Agradezco mucho el doctor J. Antonio Perez por su atenta invitaci\'on
a presentar platica en el Encuentro de Investigaci\'on. Expreso mi
reconocimiento a auditorium en muchos simposios en M\'exico donde
yo he presentado esas ideas. Reconozco la ayuda en la ortograf\'{\i}a
espa\~nola del Sr. Milton Mu\~noz Navia.

\end{document}